\documentclass[aps,prl,twocolumn,showpacs]{revtex4}  

\usepackage{latexsym}
\usepackage{graphicx}

\begin{document} 

\bibliographystyle{apsrev}

\title{Exact Results for the Crossover from Gaussian to Non-Gaussian
  Order Parameter Fluctuations in Quasi One-Dimensional Electronic Systems
} 

\author{Hartmut Monien$^*$} 

\affiliation{Yukawa Institute for Theoretical
  Physics, Kyoto University, Kyoto 606-8502, Japan
}

\received{23. February 2001}

\begin{abstract} 
  The physics of quasi one-dimensional Peierls systems is dominated by
  order parameter fluctuations. We present an algorithm which allows
  for the first time to exactly calculate physical properties of the
  electrons gas coupled to classical order parameter fluctuations.
  The whole range from the Gaussian regime dominated by amplitude
  fluctuations to the non-Gaussian regime dominated by phase
  fluctuations is accessible. Our results provide insight into the
  'pseudogap' phenomenon occurring in underdoped high-$T_{c}$
  superconductors, quasi one-dimensional organic conductors and liquid
  metals.
\end{abstract}
\maketitle

Pseudogaps have been discussed recently in the context of the high temperature
superconductors. Methods which were originally designed for studying the quasi
one-dimensional problems~\cite{LeeRiceAnderson73,Sadovski74} have been used to
describe the physics of the suppression of electronic density of states due to
antiferromagnetic or superconducting
fluctuations\cite{Chandra89,McKenzie95,Shannon00,Shannon96}. Here we present a
method which allows to describe the crossover from the Gaussian regime
dominated by amplitude fluctuations to the non-Gaussian regime dominated by
phase fluctuations without any approximation. We solve the problem originally
posed by Lee, Rice and Anderson~\cite{LeeRiceAnderson73} in 1973 exactly.

Although the pseudogap problem is more general, it first appeared in the
context of the charge density wave (CDW) systems.  The Peierls transition of
quasi one-dimensional electronic systems like K$_{0.3}$MoO$_3$ is due to the
coupling of a particular phonon mode to the electrons.  This coupling would
lead to a mean field phase transition at some temperature $T_c^{MF}$.
However, the phase transition of the one-dimensional system to the charge
density wave phase, which would break a continuous symmetry is prevented by
order parameter fluctuations. Only at some lower temperature, $T_c^{3D}$,
determined by the three-dimensional coupling of the one-dimensional systems
the charge density wave phase is established. The Kohn anomaly leads to a
softening of the phonon so that at some temperature close to the mean field
Peierls transition temperature, $T_c^{MF}$ the phonon mode can be viewed as a
static lattice distortion. The properties of the electrons in the intermediate
temperature regime $T_c^{MF} \gg T \gg T_c^{3D}$ are determined by the
coupling to 1D CDW order parameter fluctuations, which the mean-field theory
does not describe even qualitatively. All attempts to describe the electronic
properties in this regime starting with Lee, Rice and
Anderson~\cite{LeeRiceAnderson73,Sadovski74,Brasovski76,%
  OvchinikovErikhman77,McKenzie96} assume Gaussian order parameter
amplitude fluctuations. Recent
calculations~\cite{Tchernyshyov99,MillisMonien00,BartoschKopietzPRB99}
corrected a technical mistake in the original paper by
Sadovskii~\cite{Sadovski74} but confirmed that the density of states,
$N(\epsilon)$, of the electrons behaves like $N(\epsilon) \sim
\epsilon^2$ below the mean field gap for large correlation lengths in
the Gaussian model which is {\em completely unphysical}. On the other
hand, even a modest suppression of the density of states, i.e. a {\em
  pseudogap}, requires an enormous correlation length in a Gaussian
model~\cite{MillisMonien00,BartoschKopietzPRB99}.  Models taking into
account phase fluctuations only~\cite{BartoschKopietzPRB00}, which
should contain the right physics far below $T_c^{MF}$, tend to
overestimate the suppression of the electronic density at the Fermi
surface and cannot describe the physics above the mean field
transition. Thus a more sophisticated approach is needed.

We begin by defining the problem.  
The dispersion of the electrons close to the Fermi energy can be
assumed to be linear.  The Hamiltonian of the electrons has the form:
\begin{equation}
  \label{eq:Hamiltonian}
  \hat H = 
  -i v_F (R^\dagger \partial_x R - L^\dagger \partial_x L)
  + \Delta(x)   R^\dagger L
  + \Delta^*(x) L^\dagger R,
\end{equation}
where the operators $R^\dagger$ and $L^\dagger$ create left and right
moving electrons respectively.  The classical order parameter field
$\Delta(x)$ is determined by a Ginzburg Landau action given below,
$v_F$ is the Fermi velocity.  Contrary to the assumption in previous
work, it is {\em not} sufficient to describe the order parameter
fluctuations by the {\em variance} and {\em correlation length} only,
but that one needs to consider higher moments of the order parameter
correlator.  This becomes intuitively clear if one considers two
cases.  If the order parameter varies smoothly, as in the Gaussian
regime, regions where the order parameter is suppressed are smeared
out over the correlation length.  The electronic wavefunction is
spread out over a length comparable with the correlation length. The
kinetic energy is low and consequently many states can be found at low
energy even when the correlation length is large. On the other hand,
if the order parameter is established and only suppressed over a
length scale much shorter than the correlation length of the
potential, as in the non-Gaussian regime, the electronic wavefunction
decays over a distance $v_F/\Delta$ and has a large kinetic energy.
For the {\em same} correlation length and variance of the order
parameter the electronic wavefunction is much stronger suppressed for
non-Gaussian fluctuations.

Next we consider the order parameter fluctuations. For commensurate
fluctuations, the low energy electronic density of states is dominated
by the Dyson singularity which only exists in one
dimension~\cite{MillisMonien00,BartoschKopietzPRB99}. For the more
general case, the order parameter fluctuations are complex and the
Dyson singularity is absent. Therefore we will restrict our discussion
to complex order parameters. The {\em classical} complex order
parameter fluctuations, $\Delta(x)$ are described by the
Ginzburg-Landau functional:
\begin{equation}
  \label{eq:order_parameter_action}
  F[\Delta(x)] = \int_0^L\;dx/\xi_0
  \left( 
    c |\partial_x \Delta|^2 
    + a |\Delta|^2 
    + b |\Delta|^4
  \right)
\end{equation}
Close to the mean field phase transition $a$ varies linearly with
the temperature $a(T) = a' (T/T_c^{MF}-1)$, whereas $b$ and $c$ (and
therefore the length scale, $\xi_0 = \sqrt{c/a'}$) are nearly temperature
independent. In principle, the coefficients $a$, $b$ and
$c$ have to be determined self consistently from the electronic
properties. The 1D system is disordered above the 3D
ordering temperature, $T_c^{3D}$.  Nevertheless the action
Eq.~(\ref{eq:order_parameter_action}) has two different regimes: if
$a(T)$ is positive and large, the order parameter fluctuations
are centered around zero and basically Gaussian. For $a(T)$
negative and large the amplitude of the order parameter is given by
$\sqrt{<\Delta^2>}$ and only the phase fluctuations play a role.

Here we would like to sum over all configurations of the order
parameter with the Boltzmann weight, $\exp(-\Delta F[\Delta(x)]/k_B
T_c)$.  The technical problem is how to generate a sufficiently large
configuration of the order parameter (typically lengths of a chain: $L
\sim 10^7 \xi_0$) in the intermediate regime so that the electronic
properties can be calculated reliably. A Monte Carlo simulation of
such a large system can in principle be done (a more sophisticated
algorithm like the Wolff algorithm~\cite{Wolff89} has to be adopted to
avoid critical slowing down close to $T_c^{MF}$) but it turns out that
there is a much simpler way to perform the calculation.  The method
presented here is based on the transfer matrix formalism first used by
Scalapino, Sears and Ferrell~\cite{Scalapino72} to calculate the
thermodynamic properties of classical order parameter fluctuations in
one dimension {\em exactly}.  It is useful to write the free energy in
units where the length is measured in units of $\xi_0$, the size of
the order parameter in units of $\Delta_0=\sqrt{a'/2b}$ and the
temperature in units of $T_c^{MF}$, $\tau = T/T_c^{MF} - 1$ . The
reduced Ginzburg temperature, $\Delta\tau$, at which the fluctuations
start to dominate is $\Delta\tau = (a'^2/b k_B T_c^{MF})^{-2/3}$ (note
the factor of 2 between our definition and Ref.\cite{Scalapino72}). The
relevant physical parameters are the bare length scale, $\xi_0$, the
gap scale, $\Delta_0$, the mean field critical temperature,
$T_c^{MF}$, and the size of the fluctuation regime, $\Delta\tau$.

The transfer Hamiltonian for Eq.~(\ref{eq:order_parameter_action}) is
given in appropriate units by:
\begin{equation}
  \label{eq:transfer-hamiltonian}
  \frac{\hat H}{k_B T_c} = \sqrt{\Delta\tau}
  \left[ 
    - \frac{1}{2}\nabla^2 
    + \frac{1}{2}\frac{\tau}{\Delta\tau} |\vec\Delta|^2 
    + \frac{1}{4} |\vec\Delta|^4, 
  \right]
\end{equation}
where $\vec\Delta$ is a two dimensional vector of the real order
parameter components, $\Delta=\Delta' + i \Delta''$ and
$\vec\Delta=(\Delta', \Delta'')$. The Nabla operator is defined as
$\nabla=(\partial_{\Delta'}, \partial_{\Delta''})$. The prefactor of
the Hamiltonian, $\sqrt{\Delta\tau}$, can be absorbed in the length
scale.  This Schr\"odinger equation in imaginary time is equivalent to
a stochastic random walk~\cite{NegeleOrland88} where the groundstate
wavefunction is the {\em distribution function} of the spatial
coordinate of the Schr\"odinger equation. The ``spatial'' coordinate
of the transfer Hamiltonian is the order parameter fluctuation
$\Delta$ and the imaginary time corresponds to the spatial coordinate
along the chain $x$.  To see this, we introduce the wavefunction,
$\psi$, as a ratio $\psi(\Delta,x) = \Phi(\Delta,x)/\psi_0(\Delta,x)$
of an auxilary function $\Phi$ and $\psi_0$ the ground state wave
function of the anharmonic oscillator for given parameters $a$, $b$
and $c$. The function $\Phi$ obeys the following equation of motion:
\begin{equation}
  \label{eq:equation_of_motion}
  \frac{\partial\Phi}{\partial x} = 
  - \frac{1}{2}\nabla^2\Phi 
  + \nabla\cdot\left(\frac{\nabla \psi_0}{\psi_0}\;\Phi\right).
\end{equation}
This is nothing but a diffusion equation for $\Phi$. The diffusion of
the order parameter, $\Delta$, can thus be described by a Langevin
equation:
\begin{equation}
  \label{eq:Langevin}
  \frac{\partial\vec\Delta}{\partial x} = 
  - \frac{\nabla \psi_0}{\psi_0}
  + \vec\eta,
\end{equation}
where $\vec \eta = (\eta', \eta'')$, is uncorrelated Gaussian noise
in the complex plane which can be generated easily.
                                                                  
The Langevin equation Eq.~(\ref{eq:Langevin}) is nonlinear but can be
simulated very easily (see e.g.~\cite{Honerkamp93}). The distribution of
the order parameter fluctuations $\Delta$ is given by construction by
the ground state wave function of the transfer Hamiltonian. The key
observation now is that the order parameter fluctuations which can be
{\em locally} generated from the Langevin equation Eq.
(\ref{eq:Langevin}) have precisely the statistics given by the action
Eq.~(\ref{eq:order_parameter_action}). The Langevin equation, Eq.
(\ref{eq:Langevin}) can be viewed as an extremely efficient way to
generate typical order parameter fluctuations.

The method presented here is related to the path integral Monte Carlo
algorithm in which the solution of the Schr\"odinger equation is
obtained by simulating the kinetic energy with a diffusion equation
and the potential energy using a von Neumann rejection to implement
importance sampling~\cite{NegeleOrland88}.  Because the ``guiding
function'', $\psi_0$, is the solution of the transfer Hamiltonian
Schr\"odinger equation ``paths'' (configurations) in the order
parameter space are generated according to their weight in the
partition function. In some sense we have inverted the path integral
Monte Carlo method to generate the paths according to their statistical
weight. 

\begin{figure}[ht]
  \begin{center}
    \includegraphics*[width=7cm]{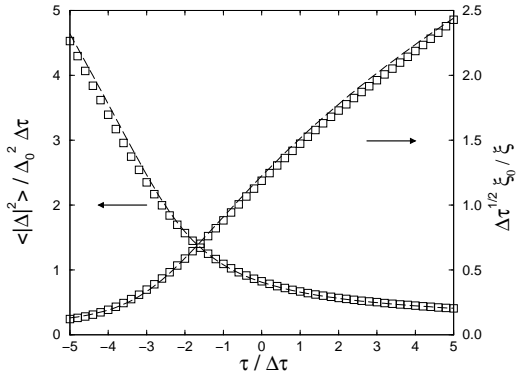}
  \end{center}
  \caption{Variance and correlation length of the order parameter 
    fluctuations as a function of reduced temperature,
    $\tau$ generated using Eq.~(\ref{eq:Langevin})
    (marked by $\Box$) for a finite system with length $L=10^6$
    compared to to the exact results by Scalapino et al.
    \cite{Scalapino72} (dashed line). }
  \label{fig:order-parameter}
\end{figure}

As a first test, we use Eq.~(\ref{eq:Langevin}) to generate a
configuration of a chain with length $L = 10^6$ and calculate the
average of the square of the order parameter fluctuation,
$<|\Delta|^2>$, and the correlation length $\xi$ and compare
to the exact results by Scalapino et al.~\cite{Scalapino72} in
Fig.~\ref{fig:order-parameter}.  It is obvious that even for a
relative short chain the calculated variance and correlation length
are basically identical to the exact results. The differences are due
to the numerical evaluation of the drift function
$\nabla\psi_0/\psi_0$. We solve the Schr\"odinger equation
Eq.~(\ref{eq:Hamiltonian}) numerically and approximate the logarithmic
derivative of the ground state wave function by a fourth order polynom
in $|\vec\Delta|$ - which apparently is a very good representation of
$|\nabla\psi_0|/\psi_0(|\vec\Delta|)$. As an aside we remark that for
the case of Gaussian order parameter fluctuations the drift term
$|\nabla\psi_0|/\psi_0(|\vec\Delta|)$ is linear and corresponds to the
Ornstein-Uhlenbeck process considered in
Ref.~\cite{BartoschKopietzPRB99}.

\begin{figure}[ht]
  \begin{center}
    \includegraphics*[width=6.5cm]{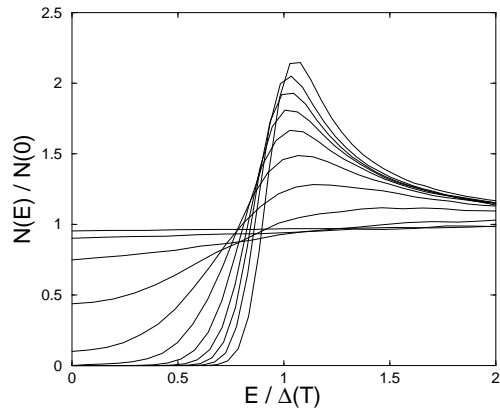}
  \end{center}
  \caption{The density of states normalized to the bare density of states,
    $N(0)$, as a function of energy, $E /\Delta(T)$ and reduced
    temperature $\tau$. The temperature varies from $\tau/\Delta\tau =
    -10 \dots 0$ in steps of one. As the temperature is decreased
    below the mean field transition the density of states is strongly
    suppressed.}
  \label{fig:dos}
\end{figure}

The density of states of Eq.~(\ref{eq:Hamiltonian}) for a given order
parameter configuration can be calculated by various methods, for
example by using a lattice version of the Hamiltonian
Eq.~(\ref{eq:Hamiltonian}) and exact
diagonalization~\cite{MillisMonien00} or some more sophisticated
method based on the phase formalism which has been developed recently
in Ref.~\cite{BartoschKopietzPRB99} (for a detailed description
see~\cite{Bartosch01}).  We calculate the density of states with the
Langevin equation for the order parameter, Eq.~(\ref{eq:Langevin})
which contains {\em amplitude} and {\em phase fluctuations}.  In this
way only {\em local} information is needed to calculate the properties
of the electrons. The requirement for storage is minimal compared to a
full simulation of the classical field.  The density of states can be
obtained by differentiating numerically the integrated density of
states. The resulting density of states as a function of energy is
shown in Fig.~\ref{fig:dos}. The energy scale at a fixed temperature
is $\Delta(T) = \sqrt{<|\Delta|^2>}$.

The calculated density of states shown in Fig.~\ref{fig:dos} has several
features which apparently resolve some of the problems of the approximations
made in previous calculations.  One of the problems of the Gaussian
approximations is that only for very large correlation lengths of the order
parameter fluctuations a pseudogap appears. Here we see that already for
relatively modest correlation lengths much above the mean field transition the
density of states at zero frequency is strongly suppressed. Our calculation
smoothly interpolates between the amplitude fluctuation dominated regime and
the phase fluctuation dominated regime.

\begin{figure}[ht]
  \begin{center}
    \includegraphics*[width=6.5cm]{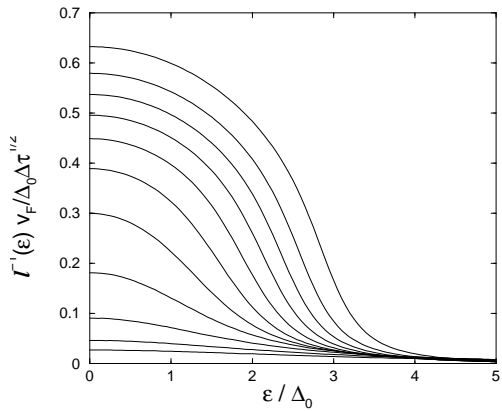}
  \end{center}
  \caption{The inverse localization length as a function of frequency for 
    reduced temperature. $\tau$ varies from $\tau/\Delta\tau = -10 \dots
    0$ in steps of one. The localization length increases with
    decreasing temperature. The lower curve corresponds to the highest
    temperature and the topmost to the lowest temperature.}
  \label{fig:localization-length}
\end{figure}

The localization length can be calculated from the Thouless
relation~\cite{Bartosch01},
and is presented in Fig.~\ref{fig:localization-length}. For
increasing temperatures the localization length decreases and
approaches zero uniformly.  
The localization length at a given energy is a
monotonic function of temperature which is different from the phase
fluctuation only model~\cite{BartoschKopietzPRB00}. This might be due
to the fact that in those models the variation of the gap scale as a
function of temperature is not taken into account.

Finally we discuss the implication of our calculation for experiments.
As an example, we calculate the temperature dependence of the
electronic spin susceptibility as a function of temperature.  The
relevant parameters are the size of the gap fluctuations $\Delta_0$
and the size of the fluctuation regime, $\Delta\tau$.
\begin{figure}[ht]
  \begin{center}
    \includegraphics*[width=6.5cm]{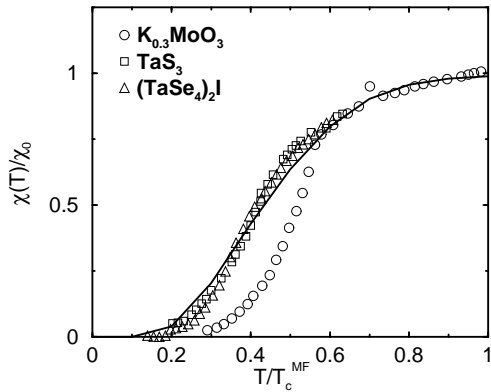}
  \end{center}
  \caption{Experimentally determined static spin susceptibilities $\chi_0(T)$
    for a number of materials from Ref.~\cite{Johnston85} and a
    tentative fit (solid line) using the calculated density of states.
    The parameters are $\Delta_0=2 T_c^{MF}$ and $\Delta\tau=0.1$.  
    }
  \label{fig:chi}
\end{figure}
Apparently excellent agreement can be achieved with the
experimentally determined Ginzburg temperature 20K for
K$_{0.3}$MoO$_3$ \cite{GruenerBook94}. At a lower temperature,
$T_{3D}$, three-dimensional ordering sets in ($T_{3D}/T_{MF}$ = 0.4,
0.6 and 0.26 for TaS$_3$, K$_{0.3}$MoO$_3$ and (TaSe$_4$)$_2$I
respectively). The departure from the purely 1D behavior can be most
clearly noted for K$_{0.3}$MoO$_3$.

To summarize, we have developed a new algorithm for studying the
electronic properties of the quasi-one-dimensional Peierls systems
which allowed for the first time to study the electronic properties in
crossover regime from the Gaussian to the non-Gaussian regime of the
order parameter fluctuations without further approximations.
Similar problems appear in the study of the pseudogap phenomenon in
the High Temperature Superconductors. More general the question of
what happens to an electronic system coupled to {\em soft} classical
degrees of freedom is relevant not only for the pseudogap but also for
the question of how to describe the moment formation in an itinerant
electronic system. These questions have been discussed in terms of
sophisticated perturbation theories, like the Parquet
approximation~\cite{Bickers89}, or self consistent approaches like the
SCR theory by Moriya~\cite{Moriya85}.  Our calculation basically
demonstrates that it is not possible to describe the moment
formation by Gaussian fluctuations (or perturbation theory), but that a
self consistent theory which takes into account the non-Gaussian
fluctuations into account is necessary.

I would like to thank M.~Imada, N.~Nagaosa, A.~Furusaki, A.~Rosch,
A.~J.~Millis, P.~Curthy, H.~Rieger, P.~Kopietz, L.~Bartosch and A.~
Sudb{\o} for useful discussions. Support from the Japanese Society for
Promotion of Science and NATO grant No. CGR~960680 is gratefully
acknowledged.

$^*$ permanent address: Physikalisches Institut, University Bonn,
D-53115 Bonn, Germany

\bibliography{pseudogap}

\begin{thebibliography}{22}
\expandafter\ifx\csname natexlab\endcsname\relax\def\natexlab#1{#1}\fi
\expandafter\ifx\csname bibnamefont\endcsname\relax
  \def\bibnamefont#1{#1}\fi
\expandafter\ifx\csname bibfnamefont\endcsname\relax
  \def\bibfnamefont#1{#1}\fi
\expandafter\ifx\csname citenamefont\endcsname\relax
  \def\citenamefont#1{#1}\fi
\expandafter\ifx\csname url\endcsname\relax
  \def\url#1{\texttt{#1}}\fi
\expandafter\ifx\csname urlprefix\endcsname\relax\def\urlprefix{URL }\fi
\providecommand{\bibinfo}[2]{#2}
\providecommand{\eprint}[2][]{\url{#2}}

\bibitem[{\citenamefont{Lee et~al.}(1973)\citenamefont{Lee, Rice, and
  Anderson}}]{LeeRiceAnderson73}
\bibinfo{author}{\bibfnamefont{P.~A.} \bibnamefont{Lee}},
  \bibinfo{author}{\bibfnamefont{T.~M.} \bibnamefont{Rice}}, \bibnamefont{and}
  \bibinfo{author}{\bibfnamefont{P.~W.} \bibnamefont{Anderson}},
  \bibinfo{journal}{Phys. Rev. Lett.}
  \textbf{\bibinfo{volume}{31}}(\bibinfo{number}{7}), \bibinfo{pages}{462}
  (\bibinfo{year}{1973}).

\bibitem[{\citenamefont{Sadovskii}(1974)}]{Sadovski74}
\bibinfo{author}{\bibfnamefont{M.~V.} \bibnamefont{Sadovskii}},
  \bibinfo{journal}{Sov. Phys. JETP}
  \textbf{\bibinfo{volume}{39}}(\bibinfo{number}{5}), \bibinfo{pages}{845}
  (\bibinfo{year}{1974}).

\bibitem[{\citenamefont{Chandra}(1989)}]{Chandra89}
\bibinfo{author}{\bibfnamefont{P.}~\bibnamefont{Chandra}},
  \bibinfo{journal}{J.~Phys.~Cond.~Matt.} \textbf{\bibinfo{volume}{1}},
  \bibinfo{pages}{10067} (\bibinfo{year}{1989}).

\bibitem[{\citenamefont{McKenzie}(1995)}]{McKenzie95}
\bibinfo{author}{\bibfnamefont{R.~H.} \bibnamefont{McKenzie}},
  \bibinfo{journal}{Phys. Rev. {\bf B}}
  \textbf{\bibinfo{volume}{52}}(\bibinfo{number}{23}), \bibinfo{pages}{16428}
  (\bibinfo{year}{1995}).

\bibitem[{\citenamefont{Shannon and Joynt}(2000)}]{Shannon00}
\bibinfo{author}{\bibfnamefont{N.}~\bibnamefont{Shannon}} \bibnamefont{and}
  \bibinfo{author}{\bibfnamefont{R.}~\bibnamefont{Joynt}},
  \bibinfo{journal}{Solid State Commun.} \textbf{\bibinfo{volume}{12}},
  \bibinfo{pages}{7405} (\bibinfo{year}{2000}).

\bibitem[{\citenamefont{Shannon and Joynt}(1996)}]{Shannon96}
\bibinfo{author}{\bibfnamefont{N.}~\bibnamefont{Shannon}} \bibnamefont{and}
  \bibinfo{author}{\bibfnamefont{R.}~\bibnamefont{Joynt}}, \bibinfo{journal}{J.
  Phys. Cond. Mat.} \textbf{\bibinfo{volume}{8}}, \bibinfo{pages}{10493}
  (\bibinfo{year}{1996}).

\bibitem[{\citenamefont{Brasovskii and Dzyaloshinskii}(1976)}]{Brasovski76}
\bibinfo{author}{\bibfnamefont{S.~A.} \bibnamefont{Brasovskii}}
  \bibnamefont{and} \bibinfo{author}{\bibfnamefont{I.~E.}
  \bibnamefont{Dzyaloshinskii}}, \bibinfo{journal}{Sov. Phys. JETP}
  \textbf{\bibinfo{volume}{44}}, \bibinfo{pages}{1233} (\bibinfo{year}{1976}).

\bibitem[{\citenamefont{Ovchinnikov and
  \'Erikhman}(1977)}]{OvchinikovErikhman77}
\bibinfo{author}{\bibfnamefont{A.}~\bibnamefont{Ovchinnikov}} \bibnamefont{and}
  \bibinfo{author}{\bibfnamefont{N.}~\bibnamefont{\'Erikhman}},
  \bibinfo{journal}{Sov.Phys.JETP}
  \textbf{\bibinfo{volume}{46}}(\bibinfo{number}{2}) (\bibinfo{year}{1977}).

\bibitem[{\citenamefont{McKenzie}(1996)}]{McKenzie96}
\bibinfo{author}{\bibfnamefont{R.~H.} \bibnamefont{McKenzie}},
  \bibinfo{journal}{Phys. Rev. Lett.} \textbf{\bibinfo{volume}{77}},
  \bibinfo{pages}{4804} (\bibinfo{year}{1996}).

\bibitem[{\citenamefont{Tschernyshyov}(1999)}]{Tchernyshyov99}
\bibinfo{author}{\bibfnamefont{O.}~\bibnamefont{Tschernyshyov}},
  \bibinfo{journal}{Phys. Rev. {\bf B}} \textbf{\bibinfo{volume}{59}},
  \bibinfo{pages}{1358} (\bibinfo{year}{1999}).

\bibitem[{\citenamefont{Millis and Monien}(2000)}]{MillisMonien00}
\bibinfo{author}{\bibfnamefont{A.~J.} \bibnamefont{Millis}} \bibnamefont{and}
  \bibinfo{author}{\bibfnamefont{H.}~\bibnamefont{Monien}},
  \bibinfo{journal}{Phys. Rev. {\bf B}} \textbf{\bibinfo{volume}{61}},
  \bibinfo{pages}{12496} (\bibinfo{year}{2000}).

\bibitem[{\citenamefont{Bartosch and Kopietz}(1999)}]{BartoschKopietzPRB99}
\bibinfo{author}{\bibfnamefont{L.}~\bibnamefont{Bartosch}} \bibnamefont{and}
  \bibinfo{author}{\bibfnamefont{P.}~\bibnamefont{Kopietz}},
  \bibinfo{journal}{Phys. Rev. {\bf B}} \textbf{\bibinfo{volume}{60}},
  \bibinfo{pages}{15488} (\bibinfo{year}{1999}).

\bibitem[{\citenamefont{Bartosch and Kopietz}(2000)}]{BartoschKopietzPRB00}
\bibinfo{author}{\bibfnamefont{L.}~\bibnamefont{Bartosch}} \bibnamefont{and}
  \bibinfo{author}{\bibfnamefont{P.}~\bibnamefont{Kopietz}},
  \bibinfo{journal}{Phys. Rev. {\bf B}}
  \textbf{\bibinfo{volume}{62}}(\bibinfo{number}{24}), \bibinfo{pages}{{\bf R}
  16223} (\bibinfo{year}{2000}).

\bibitem[{\citenamefont{Wolff}(1989)}]{Wolff89}
\bibinfo{author}{\bibfnamefont{U.}~\bibnamefont{Wolff}},
  \bibinfo{journal}{Phys. Rev. Lett.}
  \textbf{\bibinfo{volume}{62}}(\bibinfo{number}{4}), \bibinfo{pages}{361}
  (\bibinfo{year}{1989}).

\bibitem[{\citenamefont{Scalapino et~al.}(1972)\citenamefont{Scalapino, Sears,
  and Ferrell}}]{Scalapino72}
\bibinfo{author}{\bibfnamefont{D.~J.} \bibnamefont{Scalapino}},
  \bibinfo{author}{\bibfnamefont{M.}~\bibnamefont{Sears}}, \bibnamefont{and}
  \bibinfo{author}{\bibfnamefont{R.~A.} \bibnamefont{Ferrell}},
  \bibinfo{journal}{Phys. Rev. {\bf B}}
  \textbf{\bibinfo{volume}{6}}(\bibinfo{number}{9}), \bibinfo{pages}{3409}
  (\bibinfo{year}{1972}).

\bibitem[{\citenamefont{Negele and Orland}(88)}]{NegeleOrland88}
\bibinfo{author}{\bibfnamefont{J.~W.} \bibnamefont{Negele}} \bibnamefont{and}
  \bibinfo{author}{\bibfnamefont{H.}~\bibnamefont{Orland}},
  \emph{\bibinfo{title}{Quantum Many-Particle Systems}}
  (\bibinfo{publisher}{Addison-Wesley}, \bibinfo{year}{88}).

\bibitem[{\citenamefont{Honerkamp}(1993)}]{Honerkamp93}
\bibinfo{author}{\bibfnamefont{J.}~\bibnamefont{Honerkamp}},
  \emph{\bibinfo{title}{Stochastic dynamical systems : concepts, numerical
  methods, data analysis}} (\bibinfo{publisher}{VCH Wiley},
  \bibinfo{address}{New York}, \bibinfo{year}{1993}).

\bibitem[{\citenamefont{Bartosch}(2001)}]{Bartosch01}
\bibinfo{author}{\bibfnamefont{L.}~\bibnamefont{Bartosch}},
  \emph{\bibinfo{title}{Fluctuation effects in disordered {Peierls} systems}}
  (\bibinfo{year}{2001}), \bibinfo{note}{accepted for publication in Ann. Phys.
  (Leipzig)}, \eprint{, cond-mat/0102160}.

\bibitem[{\citenamefont{Johnston et~al.}(1985)\citenamefont{Johnston, Maki, and
  Gr\"uner}}]{Johnston85}
\bibinfo{author}{\bibfnamefont{D.}~\bibnamefont{Johnston}},
  \bibinfo{author}{\bibfnamefont{K.}~\bibnamefont{Maki}}, \bibnamefont{and}
  \bibinfo{author}{\bibfnamefont{G.}~\bibnamefont{Gr\"uner}},
  \bibinfo{journal}{Solid State Commun.} \textbf{\bibinfo{volume}{53}},
  \bibinfo{pages}{5} (\bibinfo{year}{1985}).

\bibitem[{\citenamefont{Gr{\"u}ner}(1994)}]{GruenerBook94}
\bibinfo{author}{\bibfnamefont{G.}~\bibnamefont{Gr{\"u}ner}},
  \emph{\bibinfo{title}{Density Waves in Solids}} (\bibinfo{publisher}{Addison
  Wesley}, \bibinfo{address}{New York}, \bibinfo{year}{1994}).

\bibitem[{\citenamefont{Bickers and Scalapino}(1989)}]{Bickers89}
\bibinfo{author}{\bibfnamefont{N.~E.} \bibnamefont{Bickers}} \bibnamefont{and}
  \bibinfo{author}{\bibfnamefont{D.~J.} \bibnamefont{Scalapino}},
  \bibinfo{journal}{Ann. Phys. (N.Y.)} \textbf{\bibinfo{volume}{193}},
  \bibinfo{pages}{206} (\bibinfo{year}{1989}).

\bibitem[{\citenamefont{Moriya}(1985)}]{Moriya85}
\bibinfo{author}{\bibfnamefont{T.}~\bibnamefont{Moriya}},
  \emph{\bibinfo{title}{Spin fluctuations in itinerant electron magnetism}},
  vol.~\bibinfo{volume}{56} (\bibinfo{publisher}{Springer series in solid-state
  sciences}, \bibinfo{address}{Berlin, New York}, \bibinfo{year}{1985}).

\end{thebibliography}
\end{document}